# Upgrade of a low-temperature scanning tunneling microscope for electron-spin resonance

Fabian D. Natterer[*‡], François Patthey, Tobias Bilgeri, Patrick R. Forrester, Nicolas Weiss, and Harald Brune[*]

Institute of Physics, École Polytechnique Fédérale de Lausanne, CH-1015 Lausanne, Switzerland

**Electron spin resonance with a scanning tunneling microscope (ESR-STM) combines the high energy resolution of spin resonance spectroscopy with the atomic scale control and spatial resolution of STM. Here we describe the upgrade of a helium-3 STM with a 2D vector-field magnet ($B_z$ = 8.0 T, $B_x$ = 0.8 T) to an ESR-STM. The system is capable of delivering RF power to the tunnel junction at frequencies up to 30 GHz. We demonstrate magnetic field sweep ESR for the model system TiH/MgO/Ag(100) and find a magnetic moment of (1.004 ± 0.001) $\mu_B$. Our upgrade enables to toggle between a DC mode, where the STM is operated with the regular control electronics, and an ultrafast-pulsed mode that uses an arbitrary waveform generator for pump-probe spectroscopy or reading of spin-states. Both modes allow for simultaneous radiofrequency excitation, which we add via a resistive pick-off tee to the bias voltage path. The RF cabling from room temperature to the 350 mK stage has an average attenuation of 18 dB between 5 and 25 GHz. The cable segment between the 350 mK stage and the STM tip presently attenuates an additional $34^{+5}_{-3}$ dB from 10 to 26 GHz and $38^{+3}_{-2}$ dB between 20 and 30 GHz. We discuss our transmission losses and indicate ways to reduce this attenuation. We finally demonstrate how to synchronize the arrival times of RF and DC pulses coming from different paths to the STM junction, a prerequisite for future pulsed ESR experiments.**

# I. Introduction

Atomic resolution topography measurements, manipulation of single atoms, and spectroscopic characterization at the single atom and chemical bond level are the unique features of scanning tunneling microscopy (STM) and, more generally, scanning probe microscopy (SPM). The spectroscopic energy resolution of an STM is limited by the Fermi-Dirac distributions of tip and sample. This stimulated the development of dedicated low-temperature SPMs, operating well below the boiling point of liquid helium [1–5]. These advanced low-temperature STMs achieve effective electron temperatures of few dozen mK, corresponding to a spectroscopic energy resolution of several μeV, ideal for the research of superconductivity and topological materials. Note that at these temperatures, broadening by the P(E) function has to be considered [6]. Electron-spin resonance based STMs [7] push this energy resolution to the 100 neV range. Similar to nuclear magnetic and electron-spin resonance, the energy resolution of an ESR-STM does not depend on temperature, and the accessible energy range is presently only limited by the available photon energy in the tunnel junction. The excellent energy resolution combined with atomic scale control is particularly useful in the study of magnetic interactions between individual surface supported spins [8–10]. The introduction of pulsed ESR-STM would enable coherent operations on the spin-manifold and quantum gate control. So far this was not achieved since the atoms that exhibit ESR-STM signals have unfavorable ratios of Rabi rate to coherence time ($T_2$) [7,9], and could be studied only by continuous wave (cw) ESR. The exploitation of atomic clock-transitions [11,12] that increase coherence times was recently suggested for a singlet-triplet transition in a coupled Ti-dimer on MgO [10]. However, the imminent advent of pulsed ESR-STM requires the introduction of additional technical upgrades to properly combine RF and DC pulses. In anticipation of adatom systems that may exhibit long enough $T_2$ or fast enough Rabi rates, we present in this paper also the required upgrade for pulsed ESR, although we don't apply it.

In this work we describe the upgrade of a $^3$He low-temperature 2D vector field STM ($B_z$ = 8.0 T, $B_x$ = 0.8 T) for ESR measurements. We first review the pre-upgrade state of the system, we then discuss the cabling (RF and DC) in the vacuum parts of the system, and we discuss our ambient cabling schematics. Following the technical description, we characterize the RF transmission at room temperature and *in-situ* by the rectifying effect of a nonlinear STM junction. We conclude by showing the ESR capability of our upgrade using the model system of TiH on MgO ultrathin films grown on Ag(100) and demonstrate how we align the arrival times of RF and DC pulses as a preparation for upcoming pulsed ESR experiments. We expect that our upgrade can easily be extended to other STMs.

## II. The existing STM

Our existing low-temperature STM [13] operates at temperatures down to 350 mK, the cooling being provided by a single-shot helium-3 cryostat (Janis, STM-UHV-He3), equipped with a superconducting 8 T out-of-plane solenoid (1914.1 Gauss/A), and a 0.8 T in-plane split-pair magnet (227 Gauss/A). The STM has four temperature modes: (1) a single-shot mode reaching 350 mK with a hold time of ~12 hours, (2) a single-shot mode that begins at 1.5 K and warms to 2.5 K over the course of 20 hours, (3) a 4.3 K mode, and (4) a 48 K mode. In order to reinitialize mode 1, the charcoal pump needs to be heated to above 30 K, which warms the STM to beyond 20 K and releases hydrogen, contaminating the sample [14]. Mode 2, on the other hand can be refreshed without exceeding 4.3 K, which is why we typically measure in that mode.

The STM is rigidly attached to the inner cryostat (insert), which is mounted onto a vertical translator. Thereby, the STM can be lifted into the position located at the center of the solenoid for measurements requiring a magnetic field, or lowered and thereby exit the solenoid for sample transfer and deposition of adsorbates. An opening in the two concentric radiation shields, the outer at solid $N_2$ and the inner at liquid He temperature, permits *in-situ* deposition of atoms using a commercial e-beam evaporator (Focus, EFM-3T).

Samples are prepared in a second ultra-high vacuum (UHV) chamber separated from the one of the cryostat by a gate valve. That chamber is equipped with typical sample preparation facilities: a sputter gun, effusion cell, e-beam evaporator, leak valves for gas dosing, and a fast-entry load lock. The samples can be cooled and heated in the manipulator to temperatures between 40 and 2000 K. Samples are transferred with this manipulator into the STM.

# III. ESR and Electrical Pump-Probe Upgrade

## A. UHV Cabling (RF and DC)

In this section, we describe the vacuum cabling from the UHV interface to the tip of the STM for the RF and DC lines. We discuss our choice of materials and outline our thoughts concerning RF attenuation and thermal anchoring. Figure 1 shows the cabling schematics across the bath-cryostat and a photo of the final assembly of the cabling at the STM level. Note that before this upgrade, the bias voltage $V_b$ was applied to the sample, but in view of the anticipated RF transmission losses through multiple point-contacts in the sample-holder and sample receptacle, we decided to invert our wiring and connect the RF line carrying the bias voltage to the tip. The sample is thus connected to virtual ground of the transimpedance amplifier (Femto, DLPCA-200).

We replace the flexible stainless steel coaxial cables that had been fed via the same cable conduit through the cryostat and that were formerly used for tunnel current ($I_t$), z-piezo (z), and bias voltage ($V_b$), by semi-rigid (SR) coaxial cables according to the wiring-schematics shown in Figure 1a. The ambient to UHV interface for the SR cables is a 4-port SMA RF feedthrough (Allectra, 242-SMAD18G-C40-4). In order to achieve acceptable RF transmission while keeping the thermal load on the cryostat minimal, we chose a stainless steel SR coax line with a silver coated center conductor for $V_b$ (Coax Co., SC-119/50-SSS-SS: 11.6 dB/m at 20 GHz and 300 K). For $I_t$, and $z$, stainless steel SR coax lines (Coax Co., SC-119/50-SS-SS: 33.3 dB/m at 20 GHz and 300 K) are sufficient because these lines carry low-bandwidth signals. The latter two lines are SR only down to the 4.3 K stage due to space requirements of attenuators and SubMiniature A (SMA) connectors.

We equipped the ~1.5 m SR coax lines with non-magnetic SMA connectors at both ends (Coaxicom, gold coated CuBe) using $Sn_{96}Ag_4$ solder and a relatively aggressive flux agent (a custom mixture of watery hydrochloric acid and zinc chloride) to remove the passivation layer. We re-soldered fresh connectors whenever the resistance between the center conductor and the shield was smaller than 1 GΩ. We then buffered the flux with a watery alkaline solution (Merck, Extran AP 13), further cleaned the solder-joints with isopropyl alcohol, methanol, and finally pumped each cable-end and its connectors with a turbo pumping station to a pressure better than $5 \times 10^{-7}$ mbar to remove residual moisture. Note that we could only add connectors onto one side of the SR lines prior to feeding the cables through the 3.7 mm diameter cable conduit of the helium-insert. We subsequently soldered SMA connectors to the open SR cable end and used sacrificial rubber plugs that we had slipped onto the cable as a seal for the pumping/drying routine explained above. The soldering of the latter connectors was rather challenging and resulted in a 1 in 5 yield only.

We bridged the stretch from 4 K to the mK-stage for $V_b$ by a ~50 cm SR cable that we wound into non-touching loops to accommodate the overhead length (see Figure 1b), which we require to establish a large enough temperature gradient. The $I_t$ and $z$ signals were connected (Huber&Suhner, SMC series) to the mK-stage via flexible stainless steel cables (Lakeshore, CC-SS). The final part of $V_b$ wiring to the tip of the STM was realized with a ~15 cm flexible silver coated copper coax line (Elspec, Gefl AWG 38) which was equipped with an angled SMA connector to allow ex-situ tip changes. We glued a Cr-tip and the flexible coax line to a homebuilt Macor tip holder using conductive silver epoxy (Epotek, H27D). Neither the coarse motion in the *x, y, z* directions, nor the scanner tube are negatively affected by the coaxial cable due to the large loop radius representing a connection of comparably low stiffness.

Having the finite cooling budget of our cryostat in mind, and noting that the heat transfer of the RF line is dominated by the micron thick Ag coating of the center conductor, we thermalized all SR cables using 0 dB attenuators ($V_b$: XMA Corp. 2782; $I_t$ and *z*: Pasternack, PE7091-0) at the 4 K and mK stage. We also coated the attenuators with vacuum grease (M&I Materials, Apiezon N) and tightly fit them into aluminum (4 K) and copper (mK-stage) holders to improve the clamping force at low temperature due to the larger thermal expansion coefficient of Al and Cu vs stainless steel. Note that our cryostat chamber is not baked in order to protect the magnetic field coils; therefore, the use of vacuum grease does not lead to outgassing. To accommodate the strain from thermal contraction, we wound the SR lines into horizontal loops of ~6 cm diameter around the 4 K stage.

## B. Integration of RF and AWG into STM circuit

In this section, we describe the integration of the RF generator and arbitrary waveform generator (AWG) into the standard STM circuitry, which is sketched in Figure 2. The default position of the relay connects the bias voltage of the STM control electronics (Nanonis, RC4/SC4) to the $V_b$ line used for regular STM topographic/spectroscopic measurements and for continuous wave (cw) ESR. The sync path shown in Figure 2 synchronizes a lock-in amplifier to the internal modulation of the RF generator for ESR sweeps. For instance, we typically chop the RF output (pulse modulation) at a frequency of 887 Hz, well inside the bandwidth of the *IV*-converter, to measure the modulated tunneling current. We combine the RF power and the DC/pulse lines via a 20 dB pick-off tee (Tektronix, PSPL5370) before connecting them to the UHV feedthrough. The voltage-drop from the pick-off port (450 Ω) is negligible compared to the GΩ resistance of the tunnel junction. The RF generator output is protected by a DC block (Pasternack, PE8224) that also serves to galvanically separate the RF generator from the STM apparatus, which might be required for certain grounding schemes. Note that a regular bias tee couples the DC path through an inductor, which acts as a low-pass for fast DC pulses, incompatible with our requirements.

We feed the DC pulses from the AWG (National Instruments, USB-6366) via the relay switch to the tip of the STM as shown in Figure 2. This mode allows all-electronic pump-probe spectroscopy [15,16] and also enables pulsed ESR operation, which uses a combination of RF and DC pulses. We will describe the pulse alignment required for this mode and for which we use the TTL trigger line also shown in Figure 2 in the section below.

We have further added 40 dB low-pass RF filters on all DC and piezo scanner lines (Mini-circuits, VLFX-105+) to prevent leaking of RF power into our control electronics and to prevent RF noise from the FPGA clock polluting our DC lines. The passband of the RF filters has a negligible effect on the ~10 ns rise time of DC pulses (here limited by our AWG electronics), which we verified separately. The filter on the $V_b$ line may be exchanged to suit other passband needs.

We placed the RF generator (Keithley, N5183B) outside of the acoustically isolated lab and connected it to the STM apparatus via a 3 m flexible RF test-cable (Teledyne, 2.92 mm FlexCore). We removed the outer polymer jacket to soften the cable and clamped it at two points between rubber pads to absorb the clicks from the mechanical step attenuator that occur when RF power ranges are switched. For all other signal and piezo lines we use low-noise coaxial cables (Huber&Suhner, G_01130_HT) to prevent pickup from slipping shields (triboelectric effect).

# IV. RF Performance

## A. RF transmission

We next characterize the RF attenuation for the RF cabling for each SR coax segment using our Schottky diode rectifier (Pasternack, PE8014). *Figure 3* shows the attenuation of the $V_b$ and $I_t$ lines measured at room temperature. The total attenuation through the cables from the RF generator to the mK-stage via the ~2 m $V_b$ line is about 22 dB at 20 GHz. After cooling, we expect less attenuation due to the improved electrical conductivity at low temperature, which would represent the maximal achievable transmission for this system. Note, the attenuation of the stainless-steel SR lines $I_t$ and $z$ exceeds the 50 dB detection limit beyond 17 GHz. As mentioned above, only one side of the last segment of the RF cabling between the tip and the mK-stage is equipped with a connector, which allowed testing of its performance only *a-posteriori*, that is, by measuring RF transmission via the tunnel junction. To that end, we use the rectifying effect of a nonlinearity in the *IV*-characteristic to quantify the effective RF amplitudes in the tunnel junction, as described by Paul and co-workers [17]. Such a nonlinearity could be a spin-flip excitation [18], a vibration [19], a rotational excitation [20–22], or any sufficiently sharp feature in the *IV*-characteristic. By setting the DC bias to $V_0$, the center of the nonlinearity, while applying RF power, the junction rectifies the AC part of the current leading to a net increase in tunnel current [17]. To first order, the increase of the tunneling current is proportional to the RF amplitude allowing to calibrate the transmitted RF voltage from the measured broadening of the *IV*-characteristic [17].

Figure 4a shows the broadening of an inelastic conductance step of a Ti cluster on MgO upon RF application ($P_{RF}$ = -1 dBm, $f$ = 16.4 GHz) demonstrating RF transmission in the tunnel junction. The observed broadening of the inelastic step at ~6.7 mV corresponds to an effective RF voltage of (1.9 ± 0.6) mV, which indicates an attenuation of ~43 dB (-1 dBm corresponds to ~282 mV). We measure the frequency dependent RF transmission in the two sweeps shown in Figure 4b and c from 10 to 26 GHz at a source power of 5 dBm and from 22 to 30 GHz at 10 dBm, respectively. We find a series of local transmission maxima of 40 to 49 dB attenuation. They are separated by regions of relatively poor transmission. We document the individual frequencies exhibiting high transmission for later ESR field-sweep measurements. The average attenuation from the generator to the junction is $52^{+5}_{-3}$ dB between 10 and 26 GHz, and $56^{+3}_{-2}$ dB between 20 and 30 GHz.

From the technical specifications of the RF line and our aforementioned room temperature characterization in *Figure 3*, we attribute a combined loss of ~5 dB to the ambient RF cable, DC block, bias-tee, and RF feedthrough. In view of the effective ~22 dB attenuation coming from the SR coaxial cable connecting the 300 K and mK-stages, the overall attenuation is relatively high. The majority of the remaining 25 to 29 dB of power is lost to the tip cable and its SMA connector. This most likely originates from the 90 degree angle between the pin inside the SMA connector and the tip cable (indicated by the dashed yellow line in Figure 1b) and which is evidenced by the ~686 MHz oscillations seen in Figure 4d that match the calculated reflections of an unterminated ~15 cm long cable (FEP velocity factor ~0.7). We suspect that the subtle increase in the STM temperature (Figure 4d), when applying RF powers above a 10 dBm level, is also mostly due to that last junction. This cable/connector assembly is a candidate for replacement when the tip is changed in the future and should bring the system to within few dB to its ~22 dB attenuation potential.

## B. ESR demonstration

To test the ESR capability of our apparatus, we investigate hydrogenated Ti adatoms (TiH) on MgO/Ag(100) [9], which is a prototypical spin ½ particle with a magnetic moment of 1 $\mu_B$ [9]. Note that our ESR measurements involve a magnetic field sweep, whereas previous STM-ESR experiments were carried out at constant external magnetic field [7–9,23,24]. The method presented here has the advantage of requiring no calibration of the transfer function [7,17], which could lead to transfer function related spurious peaks [7].

The magnetic field-sweep carried out here leads to deformations of the STM body that require adjustments of the scan-piezo positions in order to remain on the same surface spot. Figure 5a reveals a ~2 nm lateral displacement after a field-sweep from $B_z = 0$ T to 550 mT and Figure 5b shows the simultaneously recorded $z$-trace indicating deformations in the $z$ direction on the order of 1.5 nm. The overall $x$, $y$, $z$-displacements may reach several nanometers. However, they are sufficiently gradual and the field increments are small compared to the values discussed above, therefore lateral and vertical drifts can readily be compensated by atom-tracking during the field-sweep. The relative position of the tip with respect to the Ti adatom apex can thereby be held constant. Our field-sweeping routine during an ESR experiment is sketched in Figure 5c. It involves atom-tracking, the sequential application of the $n$-frequencies and a magnetic field increment/decrement. Note that our routine allows choosing the number of field-increments before executing a feedback-loop and atom-tracking cycle, indicated by the dashed boxes in the schematics of Figure 5c. The related overhead time due to the two latter steps is thus short with respect to the effective averaging time at each frequency. We typically use a dwell time of 500 ms per frequency and an atom-tracking time of 1-5 s every 5-10 field-increments. As pointed out above, the RF generator is pulse-modulated at typically 887 Hz for lock-in detection throughout the ESR-sweeps.

We finally show the ESR sweeps of hydrogenated Ti (TiH) on MgO from 700 to 500 mT in Figure 6. The different Ti species are identified by their distinct apparent heights, as shown in Figure 6a, and by their characteristic conductance spectra [9]; TiH has inelastic steps at about ±80 mV, while the spectrum of Ti is devoid of clear conductance steps as seen in panel b. We monitor the ESR signal for 3 frequencies in Figure 6c and for 5 frequencies in Figure 6d. We have vertically offset the baselines according to their frequency value (increasing from bottom to top). The sweeps reveal distinct resonances at changing field values for different frequencies, mapping the Zeeman splitting between the two spin ground states. Note that some of the larger peaks of the sweep in Figure 6c-d are artifacts due to a large lock-in amplifier time-constant leading to a memory effect from the respective previous frequencies. A larger settling time for each frequency step evidently reduces these artifacts. Nonetheless, the real ESR peaks are easily identified because we know the order in which the frequencies have been applied (see sequence # in Figure 6c-d).

We fit each resonance to a Lorentzian and mark the magnetic field positions in Figure 6c-d, onto which we map linear Zeeman trends. The slope $k = \Delta f/\Delta B$ for TiH is $(28 \pm 1)$ GHz/T. The near perfect Zeeman fit to the ESR sweep in Figure 6d, tentatively labelled Ti, yields a slope of $(28.10 \pm 0.04)$ GHz/T. These two slopes yield magnetic moments of $m = k/2\mu_0 = (1.00 \pm 0.04)$ μ$_B$ and $(1.004 \pm 0.001)$ μ$_B$, respectively. The value for TiH is in excellent agreement with the measurements of Yang and co-workers [9]. Note that the magnetic field was applied mostly in-plane in Refs. [7–10,23,24], whereas the field points along the out-of-plane direction in our system. Our measurements thus confirm the textbook like spin ½ character of TiH, which is devoid of a magnetic anisotropy energy and which was similarly reported for the Kondo system Ti/Cu$_2$N [25]. Note that we verified the spectroscopic signature of TiH before and after an ESR sweep for both measurements in Figure 6c-d. While the spectrum for Figure 6c remained identical and reminiscent of TiH, we noticed a change from TiH to Ti after the ESR sweep Figure 6d. A change of atomic species during the ESR sweep is excluded, as it would lead to a distinct jump in all ESR traces. A transformation thus occurred either before the first data point of the ESR sweep or between completion of the ESR sweep and prior to the tunneling spectroscopy measurement. The magnetic moments of both ESR sweeps are identical within the error bars and thus likely stem from the same type of TiH spin. However, we cannot exclude that our ESR sweep in Figure 6d represents a measurement of pure Ti after all. Titanium on MgO was found to be a spin-1 system via tunneling spectroscopy [9] whose gas phase magnetic moment would agree with our value of 1.004 $\mu_B$. If this were the case, our data would indicate the independent conservation of orbital and spin moment quantum numbers upon adsorption on MgO.

The average full-width at half-maximum (FWHM), $\Gamma$, of the resonances for the ESR sweep in Figure 6d is (3.0 ± 0.5) mT. From the frequency-dependent RF transmission displayed in Figure 4b-c, we correlate the effective RF-drive amplitude to their respective widths, as displayed in Figure 6d. Since the Rabi flop rate, $\Omega$, is proportional to $V_{RF}$ [7], we follow the formalism of Refs. [7,23] to extract the effective phase coherence time, $T_2^*$, without requiring knowledge of $T_1$ and $\Omega$, via the relation $\Gamma = \sqrt{1 + T_1 T_2^* \Omega^2}/\pi T_2^*$. We find a zero-RF drive $\Gamma$ of (2.5 ± 0.2) mT and a corresponding coherence time of $T_2^*$ of (4.5 ± 0.4) ns. This represents a lower bound for the intrinsic $T_2$ because additional broadening mechanisms related to the vibrational stability of the tip [7], magnetic fluctuations [23], and decoherence by tunneling electrons [23] have not been accounted for.

For different ESR sweeps, the Zeeman trends show zero field intercepts in the range of -30 to +65 mT. We attribute these offsets to the tip-stray field originating from the magnetic atoms adsorbed onto the tip apex [9,26,27]. As expected, we observe that the stray field is unique for every micro-tip. Although the tip-field influences the resonance frequency, our magnetic field-sweep routine is insensitive to that effect because we use the same micro-tip for the entire sweep and we fix the tip-atom distance via atom-tracking. The magnetic field-sweep ESR described here notably allows measuring the magnetic dipole-dipole coupling between ESR active spins and nearby magnetic moments. Unlike in frequency-sweep ESR [8], the magnetic dipole field of a coupled spin would be measured directly via the shift of the resonance position without requiring prior calibration of the ESR sensor moment.

## C. RF and DC pulse alignment

Pulsed ESR measurements enable advanced insights into surface supported spins and their magnetic interactions using Rabi oscillations, Ramsey fringes, spin-echoes, and other quantum operations. These experiments rely on the exact timing between RF pulses for coherent driving and DC pulses for spin-projections. However, the propagation times of electric signals depend on cable length as well as material dependent factors, which need to be accounted for. We demonstrate here a method to calibrate and control the temporal alignment of RF and DC pulses with nanosecond precision. In order to measure their temporal alignment, we apply RF and DC voltages to a nonlinear *IV*-characteristics and exploit $I(V_{DC} + V_{RF}) \neq I(V_{DC}) + I(V_{RF})$, which is frequently used in time-resolved experiments [15,28]. We demonstrate the measurement principle by using a Schottky RF diode rectifier (Pasternack, PE8014) to emulate a nonlinear *IV*-tunnel junction.

The RF and DC pulses originate from different generators and travel along separate paths before they are combined in the pick-off tee from where they advance on the same cable to the diode (or tunnel-junction, see Figure 2). The simultaneous arrival of RF and DC pulses leads to a distinct current owing to the nonlinear *IV*-relationship and thus allows finding the trigger delay for which RF and DC pulses overlap.

In order to determine this delay, we create two DC pulse trains; one serves as the 5V-TTL reference that triggers the RF generator to initiate RF pulses. The other DC pulse train is applied to the diode and corresponds to the voltage pulses used for measuring spin-projections. We set the pulse-width of both pulses to $t_{DC}$ = 5 µs and repeat them with a period of 94.9 µs, as sketched in Figure 7. We offset the TTL and DC pulses by one-half period. The TTL trigger initiates an RF pulse of variable trigger delay $t_{delay}$ and length $t_{RF}$ = 500 ns that we want to overlap with the DC pulse. Upon sweeping $t_{delay}$, we observe a distinct change in the diode current when the DC and RF pulses start to merge/overlap, seen as a dip in the differentiated signal at about 47 µs in Figure 7. Similarly, when the overlap of DC and RF pulses ends, the signal shows a distinct peak. The position of the first dip allows us to find the delay time at which the RF pulse exactly precedes the DC pulse, which, for instance, represents the alignment scheme of RF and DC pulses in a Rabi oscillation experiment. The distance between peak and dip is exactly 5 µs, which is also the DC pulse length. Further, the widths of the dip and peak scale with the length $t_{RF}$ of the RF pulse.

In order to demonstrate the influence of different path lengths on the propagation time, we have introduced an additional 10 m cable to the trigger line to delay the arrival time of the RF executing trigger and consequently also the RF pulse (see loop in Figure 2 and inset in Figure 7). Since the TTL and the DC pulses are created at the same location in the AWG, the DC pulse gains a head start with respect to the RF pulse at the pick-off tee, which is seen as an earlier arrival (green) in the overlap signal in Figure 7. We find a delay of (45 ± 2) ns between the 10 m and 0 m trigger cases, as seen by the overlaid Gaussian fits to the respective dips in Figure 7. The effective signal propagation velocity is (0.74 ± 0.03) c, in good agreement with the cable specifications. Our pulse-alignment measurements also indicate an attractive way to permanently introduce a time-delay into the TTL trigger line to compensate for different $V_b$ and RF cable lengths without compromising the latter two signal qualities. The TTL signal level is more resilient than the typically used smaller signal amplitudes on the DC and RF side.

# V. Summary and Outlook

We have described the upgrade of an existing low-temperature (0.35 K, 8 T) STM into an ESR capable device. Our work shows RF transmission from ambient to the tip of the STM up to 30 GHz at cryogenic temperatures. We demonstrate the ESR capabilities of our upgraded apparatus on the model system TiH on MgO. We find a magnetic moment for TiH of (1.00 ± 0.04) $\mu_B$, in excellent agreement with previous work [9] and a value of (1.004 ± 0.001) $\mu_B$ for an ESR sweep using 5 monitoring frequencies and recorded above an adsorbate either representing TiH or pure Ti. Our ESR field-sweeping method provides direct insight into the magnetic field dependent Zeeman energy of the spin ground states. It should enable the calibration-free measurement of nearby magnetic stray-fields via shifts of their resonance position. We illustrate our cabling concept that allows the simultaneous application of fast DC and RF pulses, which may be used for future pulsed ESR investigations. We further demonstrate a method to align the arrival times of DC and RF pulses with nanoseconds precision, which are traveling along paths of different propagation times. We identify potential pitfalls that compromise effective RF transmission and highlight ways how they can be overcome to achieve an RF attenuation close to the 22 dB potential of our apparatus.

## Methods

The MgO layer is grown by exposing an atomically clean Ag(100) single crystal to an Mg flux from an effusion cell in an oxygen partial pressure of ~1x10$^{-6}$ mbar at a sample temperature of 773 K and using a cooling rate of 15 K min$^{-1}$ [29,30]. Iron and Ti atoms are dosed from a commercial e-beam evaporator onto the sample in the STM position. Upon deposition, some Ti adatoms are hydrogenated from the residual gas [9,14] while others remain pure Ti. The tip is etched electrochemically from 0.5 mm x 0.5 mm rods of bulk Cr following recipes described in Refs. [31,32]. However, due to several tip-indentations into the clean Ag(100) parts of the surface, our tip-apex is most likely Ag terminated. Therefore we transfer individual Fe atoms by vertical atom manipulation (-400 pm, 550 mV) from the sample to the tip in order to spin-polarize it [33]. The spin-polarization is tested on Fe by 2-state switching [34] or on Ti via asymmetric conductance steps [9].

# Acknowledgement

We are indebted to the electronic and mechanical workshops of our Institute for their expert assistance. We thank Kai Yang for discussions. F.D.N., T.B., and P.R.F. acknowledge support from the Swiss National Science Foundation under project numbers PZ00P2_167965 and 200020_176932. P.R.F. acknowledges support from the Fulbright U.S. Student Program.

*To whom correspondence should be addressed: fabian.natterer@uzh.ch and harald.brune@epfl.ch

‡Present address: University of Zurich, Winterthurstrasse 190, CH-8057 Zurich

**Figure Captions**

Figure 1 | **Schematics and view of the upgraded STM system. a**, Cabling schematics inside the vacuum chamber to the tip of the STM. The red color sketches the path of the RF line, $V_b$, which is made from a stainless steel semi-rigid (SR) coaxial line with silver-coated center conductor. The two other lines, $z$, and $I_t$, are made from a SR stainless steel coaxial line up to the 4K stage and are then continued by a flexible stainless steel coaxial cable. **b**, Photo of the STM showing the $V_b$, $z$, and $I_t$ lines. The $V_b$ line connecting the 4 K and the mK-stage is wound into a loop to reduce the heat-transmission. The dotted angle indicates the 90° SMA connector that connects the flexible coaxial cable to the tip of the STM.

Figure 2 | **Ambient side wiring schematics.** Integration of RF and arbitrary waveform generator (AWG) into the STM circuit. The pick-off tee and relay permits the application of fast DC pulses from the AWG onto the RF path for pump-probe spectroscopy or pulsed ESR experiments. The low-pass filter has a selectable passband. The trigger adjust allows compensation of different propagation times in RF and DC lines, as described in the text.

*Figure 3* | **Transmission characteristics of semi-rigid cables.** The silver coated ~2 m SR line, $V_b$, has an average room-temperature attenuation of 18 dB from 5 to 25 GHz. It bridges the thermal gradient between the ambient 300 K to the UHV mK part of the STM, as seen in the inset to the right. The weak ~220 MHz oscillations in the $V_b$ line are due to small reflections at the mK-stage and map to a cable length of ~47 cm, in agreement with the length of the segment connecting the 4 K to the mK-stage. The ~1.5 m stainless SR for $z$ and $I_t$ lines have a much larger attenuation that quickly drops below the diode detection limit around 17 GHz.

Figure 4 | **Quantifying RF transmission to the STM tip. a,** Differential conductance steps at $V_0 = 6.7$ mV of an adsorbate on MgO used for RF rectification. The application of RF power leads to a step broadening of about 2 mV for a source power of -1 dBm at 16.4 GHz. The arrow denotes $V_0$, the center of the nonlinearity, used for RF rectification. **b,** and **c,** Transmission of RF power to the STM junction at constant source power of 5 dBm and 10 dBm, respectively, measured via the rectifying effect of the nonlinearity at $V_0$. ($V_0 = 6.7$ mV, $I_t = 100$ pA, $\Delta f = 5$ MHz, $t_{dwell} = 250$ ms). **d,** Frequency dependent temperature increase of the STM body at 20 dBm source power with respect to the 4.3 K base temperature ($\Delta f = 5$ MHz, $t_{dwell} = 1$ s).

Figure 5 | **STM deformation and atom-tracking sequence during a magnetic field-sweep. a,** STM image before and after sweeping the magnetic field $B_z$ from zero to 550 mT. The sweep results in a lateral displacement of about 2 nm indicated by the apparent shift of the Fe and Ti atom positions marked by the dashed circles. **b,** The vertical tip-height is monitored during a field-sweep by tracking a Ti adatom. A permanent height-change of ~400 pm is noted. ($V_b$ = 150 mV, $I_t$ = 10 pA). **c)** Timing diagram during an ESR field-sweep showing atom-tracking and the engagement of the feedback loop. The atom is tracked at every $j^{th}$ field increment and the ESR signal is recorded for a set of frequencies that have good transmission.

Figure 6 | **ESR sweeps on TiH on MgO/Ag(100). a**, Topography of Fe, Ti, and TiH showing different apparent heights on MgO. **b,** Spectroscopy used to further identify Ti and TiH. **c,** ESR sweep on TiH, monitored at 3 frequencies, showing resonances and their corresponding Lorentzian fits superimposed. The peak positions follow a Zeeman trend of slope (28 ± 1) GHz/T indicated by the straight line. **d**, ESR sweep on Ti monitored at 5 frequencies shows a linear Zeeman trend of (28.10 ± 0.04) GHz/T. The order by which the frequencies have been applied is indicated at the right of either sweep and allows to identify spurious signals due to a memory effect of the lock-in amplifier. ($V$ = 60 mV, $I$ = 30 pA, 20 dBm RF, $T_{STM}$ ~ 2 K). **e,** Resonance width vs driving strength used to determine the effective coherence time, $T_2^*$.

Figure 7 | **RF and DC pulse alignment.** Measurement of the relative propagation times of DC and RF pulses via the rectifying effect of a nonlinearity, here demonstrated with an RF diode. A DC pulse of $t_{DC}$ = 5 µs and an RF pulse of $t_{RF}$ = 500 ns are applied at varying delays $t_{delay}$ between the RF trigger and the DC pulse. When both pulses overlap, the diode current changes, which is used to find the delay time $t_{delay}$ at which the pulse-overlap starts. Adding a cable of 10 m to the trigger line demonstrates the temporal resolution of the onset measurement, resulting in a shift of (45 ± 2) ns.

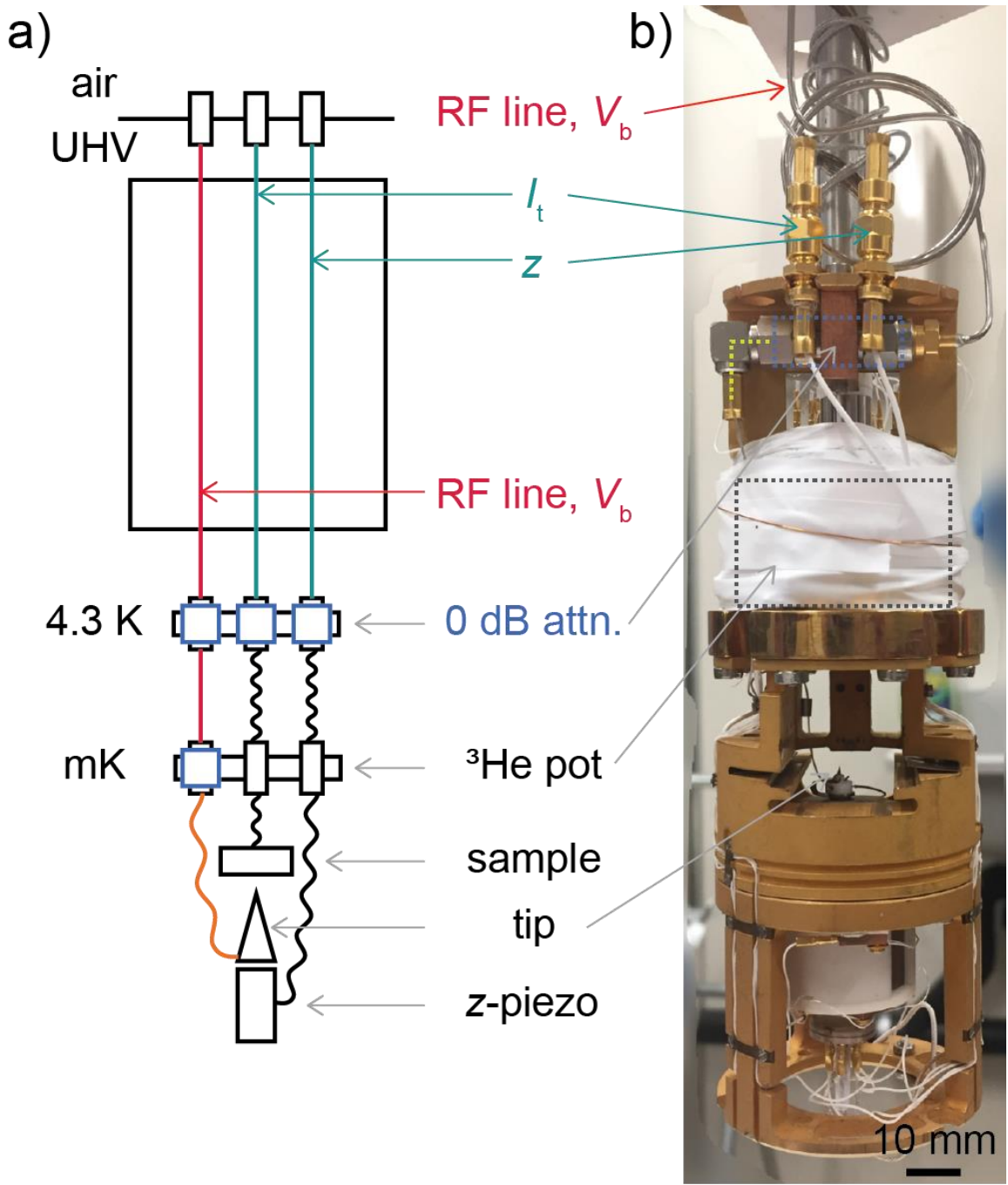

*Figure 1: STM, one column*

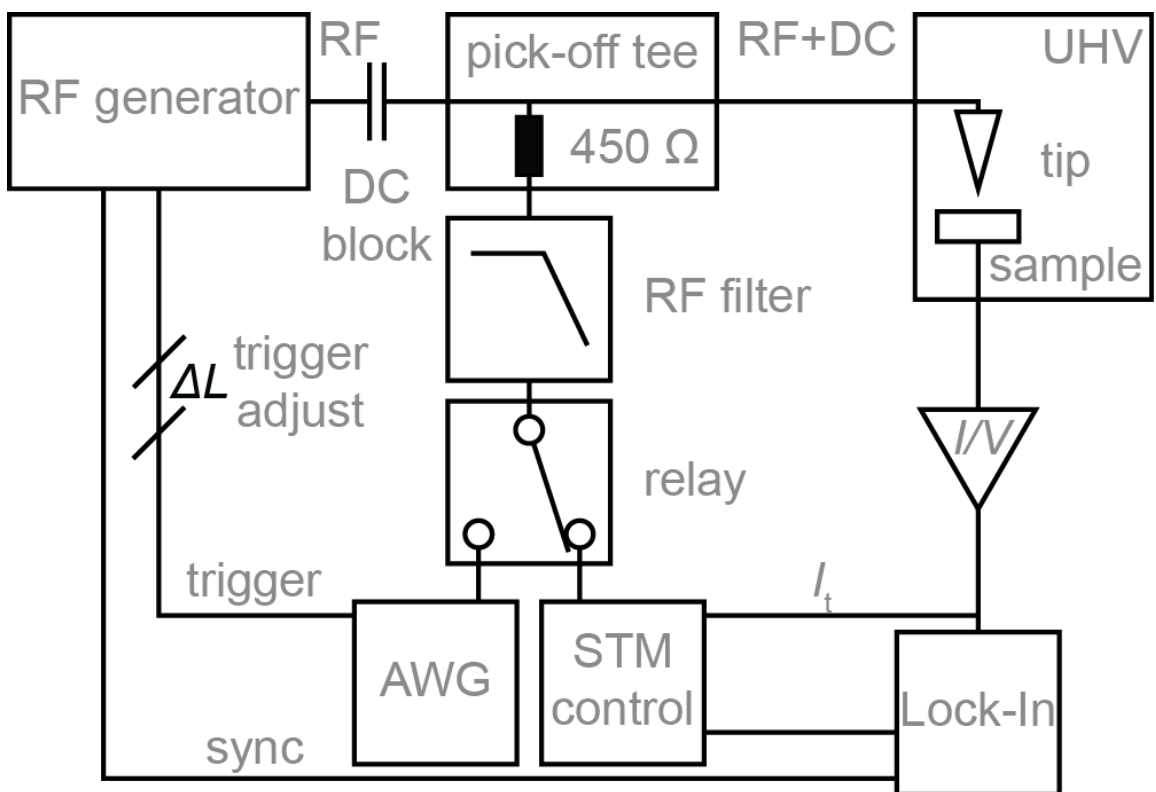

*Figure 2: AWG cabling, one column*

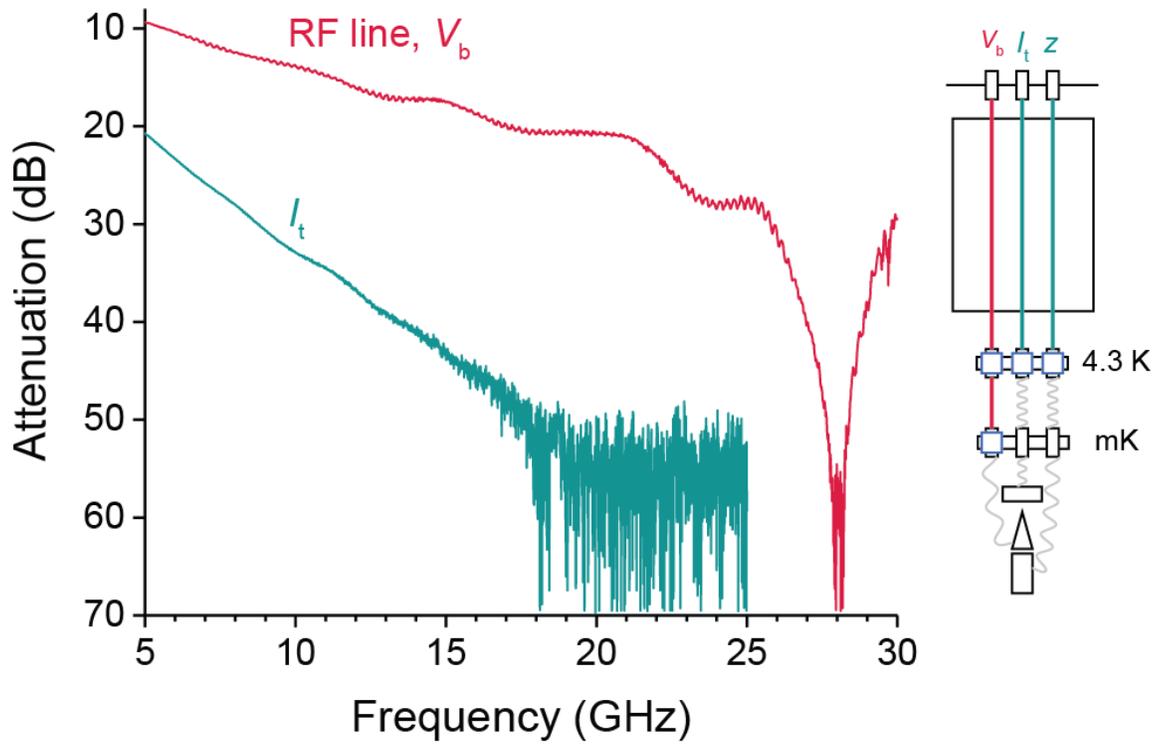

*Figure 3: RT transmission, one column*

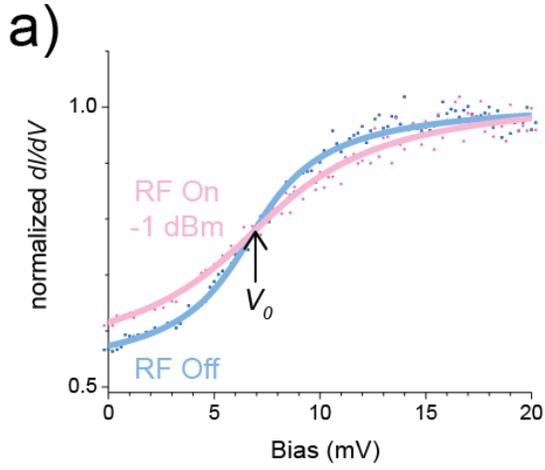
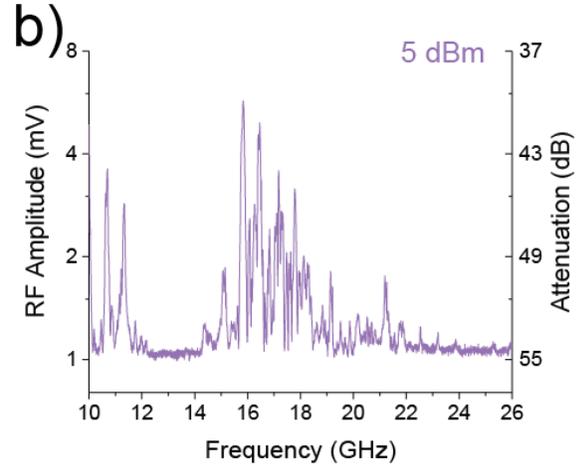
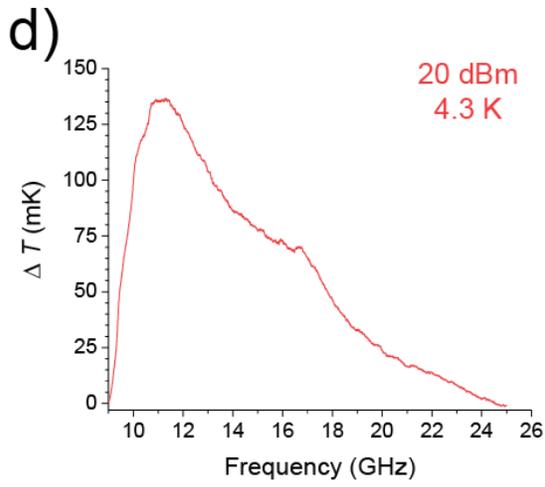
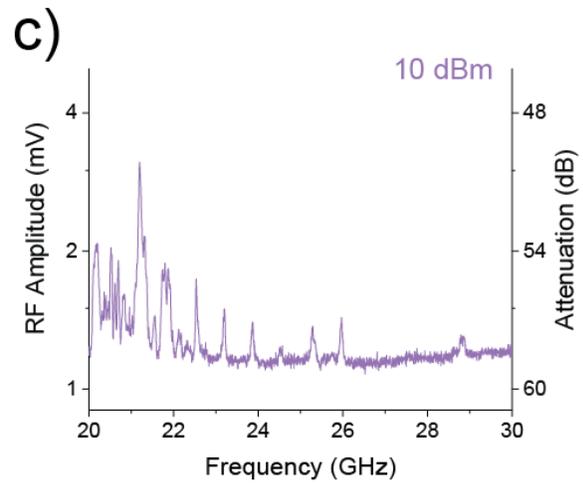

*Figure 4 rectification, one column*

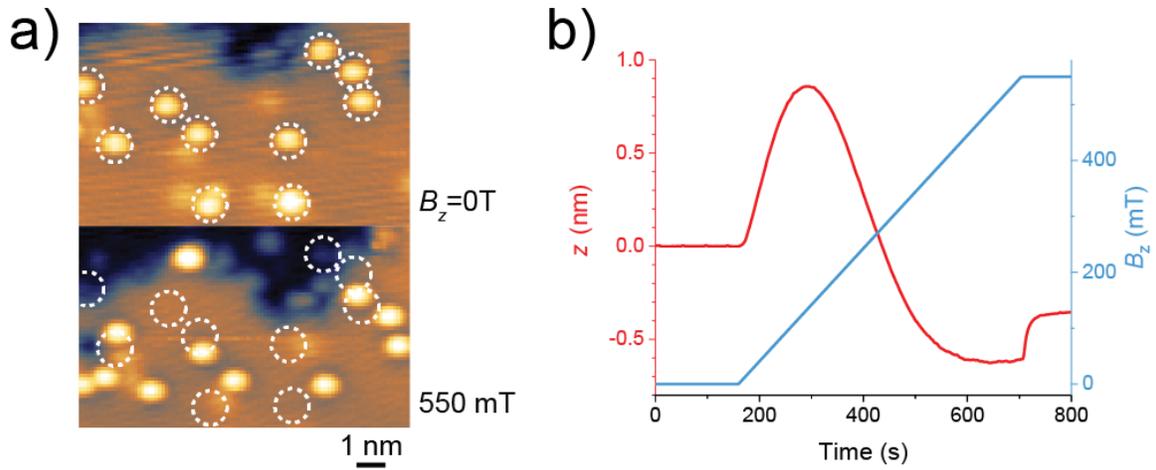
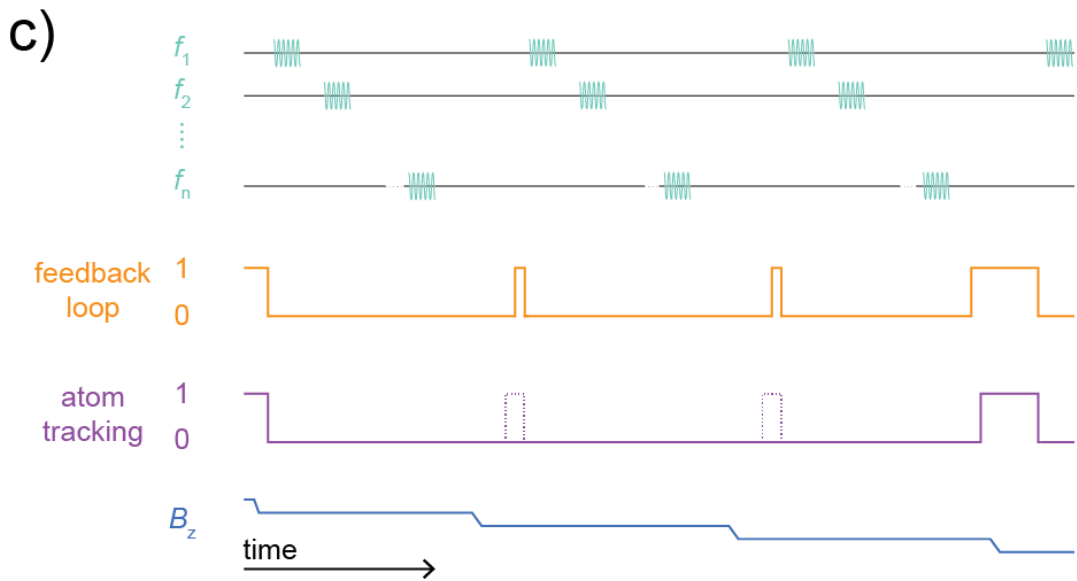

*Figure 5: field-sweep, one column*

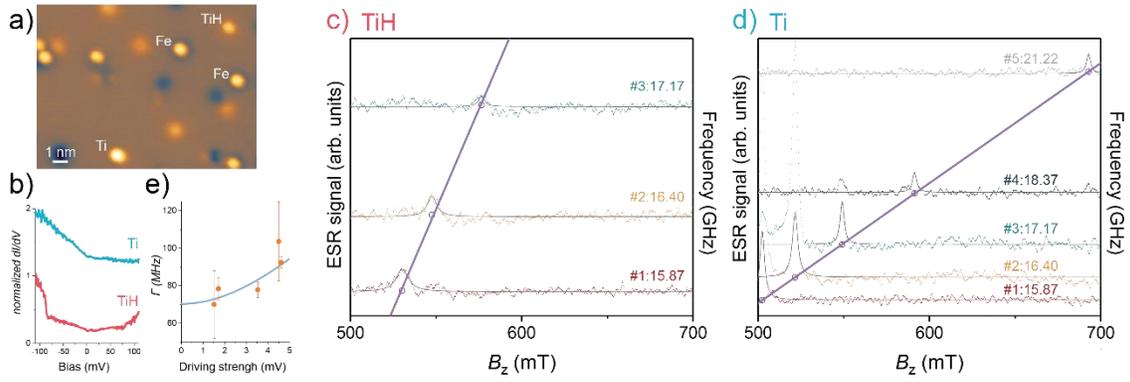

*Figure 6: ESR sweep, two columns*

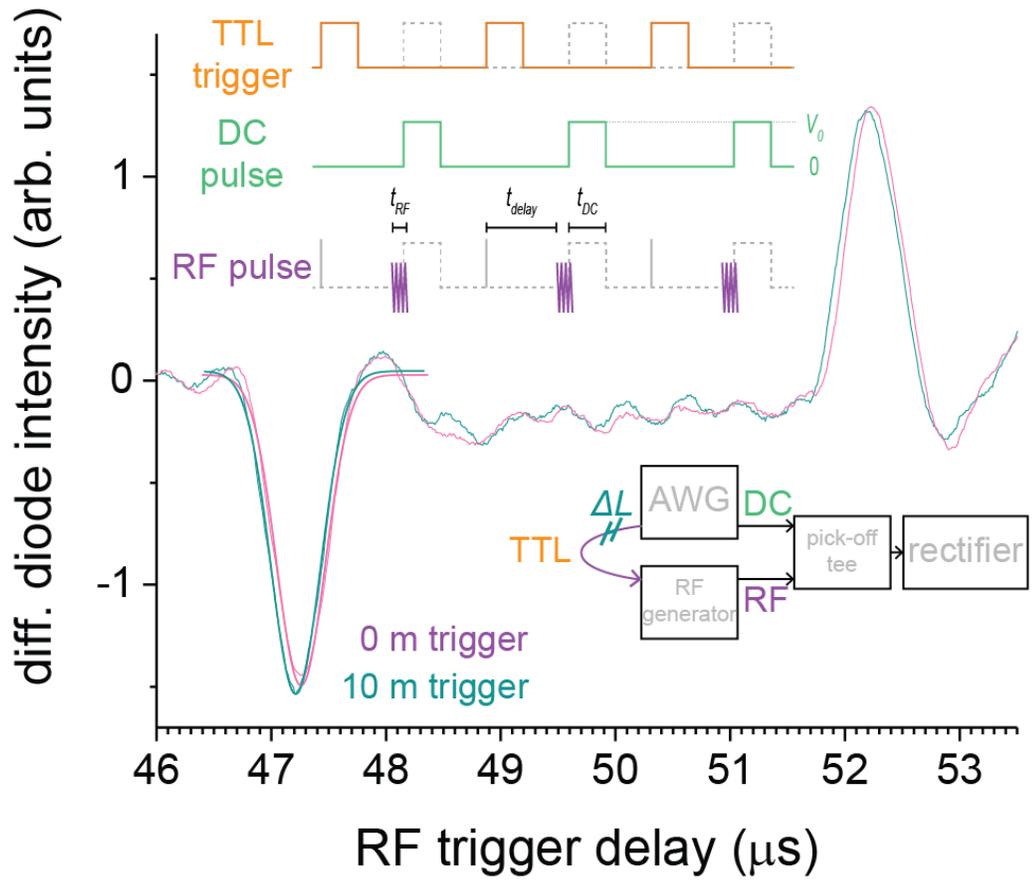

*Figure 7: Pulse Alignment, one column*